\documentclass[aps, prd,twocolumn,groupedaddress,showpacs,nofootinbib]{revtex4-1}
\pdfoutput=1

\usepackage{amssymb,amsfonts,amsmath,bm,dsfont}
\usepackage{graphicx}
\usepackage{hyperref}
\usepackage{multirow}
\usepackage{color}
\usepackage[tight]{units}

\newcommand{\fig}[1]{Fig.~\ref{#1}}
\newcommand{\Tab}[1]{Table~\ref{#1}}
\newcommand{\Sec}[1]{Sec.~\ref{#1}}
\newcommand{\Eq}[1]{Eq.~(\ref{#1})}
\newcommand{\eq}[1]{(\ref{#1})}

\newcommand{\SU}[1]{\mathrm{SU}(#1)}

\newcommand{\be}{\begin{equation}}
\newcommand{\ee}{\end{equation}}

\newcommand{\pslash}{p\hspace{-0.475em}\slash}

\newcommand{\tr}[1]{\,\text{tr}\left[#1\right]}

\renewcommand{\Re}{\mathfrak{Re}\,}

\DeclareMathOperator{\e}{e}

\begin{document}
%\sloppy

\date{\today}

\title{Lattice QCD Green's functions in maximally Abelian gauge: infrared Abelian dominance and the quark sector}
%\titlerunning{Lattice QCD Green's functions in maximally Abelian gauge}

\author{Mario~Schr\"ock}
\email{mario.schroeck@roma3.infn.it}
\affiliation{Istituto Nazionale di Fisica Nucleare (INFN), Sezione di Roma Tre, 00146 Rome, Italy}
\author{Hannes~Vogt}
\email{hannes.vogt@uni-tuebingen.de}
\affiliation{Institut f\"ur Theoretische Physik, Auf der Morgenstelle 14, 72076 T\"ubingen, Germany}

%\author{Mario~Schr\"ock\inst{1} \and Hannes~Vogt\inst{2}% etc
%}                     % Do not remove
%
%
%\institute{Istituto Nazionale di Fisica Nucleare (INFN), Sezione di Roma Tre, 00146 Rome, Italy \and Institut f\"ur Theoretische Physik, Auf der Morgenstelle 14, 72076 T\"ubingen, Germany}

\begin{abstract}
On lattice gauge field configurations with $2+1$ dynamical quark flavors, we investigate the momentum space quark and gluon propagators in the combined maximally Abelian plus $U(1)_3\times U(1)_8$ Landau gauge. We extract the gluon fields from the lattice link variables and study the diagonal and off-diagonal gluon propagators. We find that the infrared region of the transverse diagonal gluon propagator is strongly enhanced compared to the off-diagonal propagator. The Dirac operator from the Asqtad action is inverted on the diagonal and off-diagonal gluon backgrounds separately. In agreement with the hypothesis of infrared Abelian dominance, we find that the off-diagonal gluon background hardly gives rise to any non-trivial quark dynamics while the quark propagator from the diagonal gluon background closely resembles its Landau gauge counterpart.
\end{abstract}


\maketitle

%\keywords{Lattice QCD, chiral symmetry breaking, confinement, maximally Abelian gauge, infrared Abelian dominance}

\section{Introduction}
Despite the long standing acceptance of quantum chromodynamics (QCD) as the correct theory to describe the strong 
interactions of quarks and gluons, a basic understanding of its main characteristic features, the dynamical breaking of 
the chiral symmetry and confinement, is still lacking.
Several scenarios for the underlying mechanism of confinement have been suggested over the decades. Under the most popular ones are the Kugo--Ojima \cite{Kugo:1979gm} and the Gribov--Zwanziger scenarios \cite{Gribov:1977wm, Zwanziger:1992qr}, as well as the dual superconductor picture \cite{Nambu:1974zg, 'tHooft:1977hy, Mandelstam:1974pi}. Some common aspects of the different confinement criteria have been investigated in \cite{Reinhardt:2008ek, Dudal:2009xh, Suzuki:1983cg, Hata:1992np, Schaden:2013ffa}.

The dual superconductor picture of confinement is especially appealing since it offers a rather intuitive approach to confinement.
Type II superconductors in their superconducting phase are known to repel external magnetic fields below a critical 
value of the external field strength. If the magnetic field exceeds that value, tubes of magnetic flux (Abrikosov 
vortices) begin to penetrate the superconductor.
The flux tubes are encircled by Cooper pairs which squeeze the latter.
Identifying the penetrating magnetic flux tubes with the color electric field of the Yang--Mills vacuum, and moreover 
the condensed electric monopoles (Cooper pairs) with color magnetic monopoles, one finds a dual picture where 
hypothetical magnetic monopoles at the beginning and end of the Abrikosov vortices correspond to the confined quarks of 
QCD.
Therefore, it is suggested that confinement is due to the condensation of color magnetic monopoles which squeeze the 
color electric flux tube between quarks and antiquarks.
The dual superconductor picture has a far reaching consequence: the Abelian parts of the gauge fields should dominate 
the nonperturbative infrared (IR) dynamics \cite{Ezawa:1982bf}.

While confinement is the reason why individual quarks and gluons have never been observed in experiment, the theory 
still allows to investigate correlation functions of single quark and gluon entities: the QCD Green's functions. QCD is 
a gauge theory and therefore the gauge has  to be fixed in order to study the fundamental two-point functions. The 
maximally Abelian gauge, as the name suggests, rotates the gauge fields such that the diagonal, i.e., Abelian parts of 
the gauge fields are enhanced over the off-diagonal parts. This renders the maximally Abelian gauge particularly 
suitable to study IR Abelian dominance.

Lattice QCD provides an approximation to the continuum formulation with a finite number of degrees of freedom which 
allows to perform numerical simulations. Various attempts of investigating the dual superconductor picture in lattice 
QCD have been carried out, see, e.g., \cite{DelDebbio:1994sx, Carmona:2002ty}.  IR Abelian dominance has been 
demonstrated in lattice gauge field theory in SU(2) \cite{Amemiya:1998jz, Bornyakov:2003ee, Gongyo:2014lxa} and more 
recently in SU(3) \cite{Gongyo:2012jb, Gongyo:2013sha} by studying the infrared behavior of the gluon propagator. In 
the recent study \cite{Sakumichi:2014xpa} almost perfect Abelian dominance of the string tension on large physical 
volumes in quenched SU(3) has been found.
Complementary to the lattice approach, Abelian dominance has been analyzed perturbatively \cite{Quandt:1997rg, 
Quandt:1997rw} and in Dyson--Schwinger and renormalization group equation studies \cite{Huber:2009wh, Huber:2010ne, Huber:2011fw}.

It is interesting to study the influence of Abelian dominance on chiral symmetry breaking and the hadron spectrum;
in \cite{Woloshyn:1994rv} it has been found that SU(2) Abelian projected fields give a chiral condensate which closely 
resembles the results of strongly coupled gauge theory.
In \cite{Kitahara:1998sj} quenched hadron spectra in Abelian gauge fields, extracted by maximal Abelian projection have 
been studied: the ratios of the hadron mass to the square root of the string tension of the Abelian fields are similar 
to those of the full SU(3) theory. The authors concluded that they have found Abelian dominance (and monopole 
dominance) for the hadron spectra.

In the current paper we advance previous investigations of the gluon propagator in maximally Abelian gauge from pure Yang--Mills theory \cite{Gongyo:2012jb, Gongyo:2013sha} to full dynamical QCD by adopting $N_f=2+1$ gauge field configurations. Furthermore, we  analyze for the first time the maximally Abelian gauge quark propagator. In order to obtain insights in the dependence of chiral symmetry breaking and the dynamic mass generation of quarks on the type of gluon background, we invert the Dirac matrix separately on the diagonal and off-diagonal gluon fields.

The remainder of this work is structured as follows. In \Sec{sec:MAGlattice} we introduce the maximally Abelian gauge 
and discuss some aspects of its implementation on the lattice. After reviewing the methods to extract the QCD 
Green's functions on the lattice in \Sec{sec:QCDprops}, we list details of our lattice setup in \Sec{sec:results} and 
present our results.

\section{The maximally Abelian gauge on the lattice}\label{sec:MAGlattice}
The continuum gauge fields are given by
\be
  A_\mu(x) = \frac{1}{2}\sum_{i=1}^8 \lambda_i A_\mu^{(i)}(x)\,,
\ee 
where the $\lambda_i$ are the Gell-Mann matrices and $A_\mu^{(i)}(x)$ are real. 
On the lattice, the latter translate to the lattice link variables $U_\mu(x)\in \SU{3}$. The continuum and lattice fields are related via 
\be\label{eq:linkVars}
  U_\mu(x) = \e^{iag_0A_\mu(x)}
\ee
with $a$ being the lattice spacing and $g_0$ the bare coupling constant. A gauge transformation in the language of 
lattice QCD reads
\be
  U_\mu(x) \to g(x)\, U_\mu(x) \,g(x+\hat\mu)^\dagger
\ee
with local gauge transformations $g(x)\in \SU{3}$.

The maximally Abelian gauge (MAG) aims at minimizing the off-diagonal part of the gauge fields, i.e., $A_\mu^{(i)}(x)$ with $i\neq 3, 8$. This is equivalent to maximizing the following functional of the link variables,
\begin{gather}\label{eq:MAG_SU3_functional}
	F_{\mathrm{MAG3}}^g[U]=\sum_{x,\mu}
	       \tr{ \lambda_3 U_\mu(x) \lambda_3 U_\mu(x)^\dagger }\\
	      +\tr{\lambda_8 U_\mu(x) \lambda_8 U_\mu(x)^\dagger }
\end{gather}
where $\lambda_3$ and $\lambda_8$ build the Cartan subalgebra of $\SU{3}$.
Once the functional \Eq{eq:MAG_SU3_functional} resides in a local maximum, the gauge condition
\begin{gather}
	\theta = \frac{1}{VN_d}\sum_{x,j}  
	\Big( \sum_\mu 
			u^{(j)}_\mu(x)\,\sigma_3\, u^{(j)}_\mu(x)^\dagger \\
	      + u^{(j)}_\mu(x-\hat\mu)^\dagger \,\sigma_3\, u^{(j)}_\mu(x-\hat\mu) \Big)^2
\end{gather}
becomes small.
Here $N_d$ is the number of Euclidean spacetime indices, $V$ the number of lattice sites and the $u_\mu^{(j)}(x)$, 
$j=1, 2, 3$ are the SU(2) subgroup elements of the link variables $U_\mu(x)$.\footnote{Two SU(2) matrices overlap on the diagonal of $U_\mu(x)$ and the third one consists of the corners of $U_\mu(x)$.}
In practice we reach a gauge precision of $\theta<10^{-13}$.
More details of the implementation can be found in Refs.~\cite{Schrock:2012fj, Stack:2002sy}.

\begin{figure}[htb]
	\center
	\includegraphics[width=.99\columnwidth]{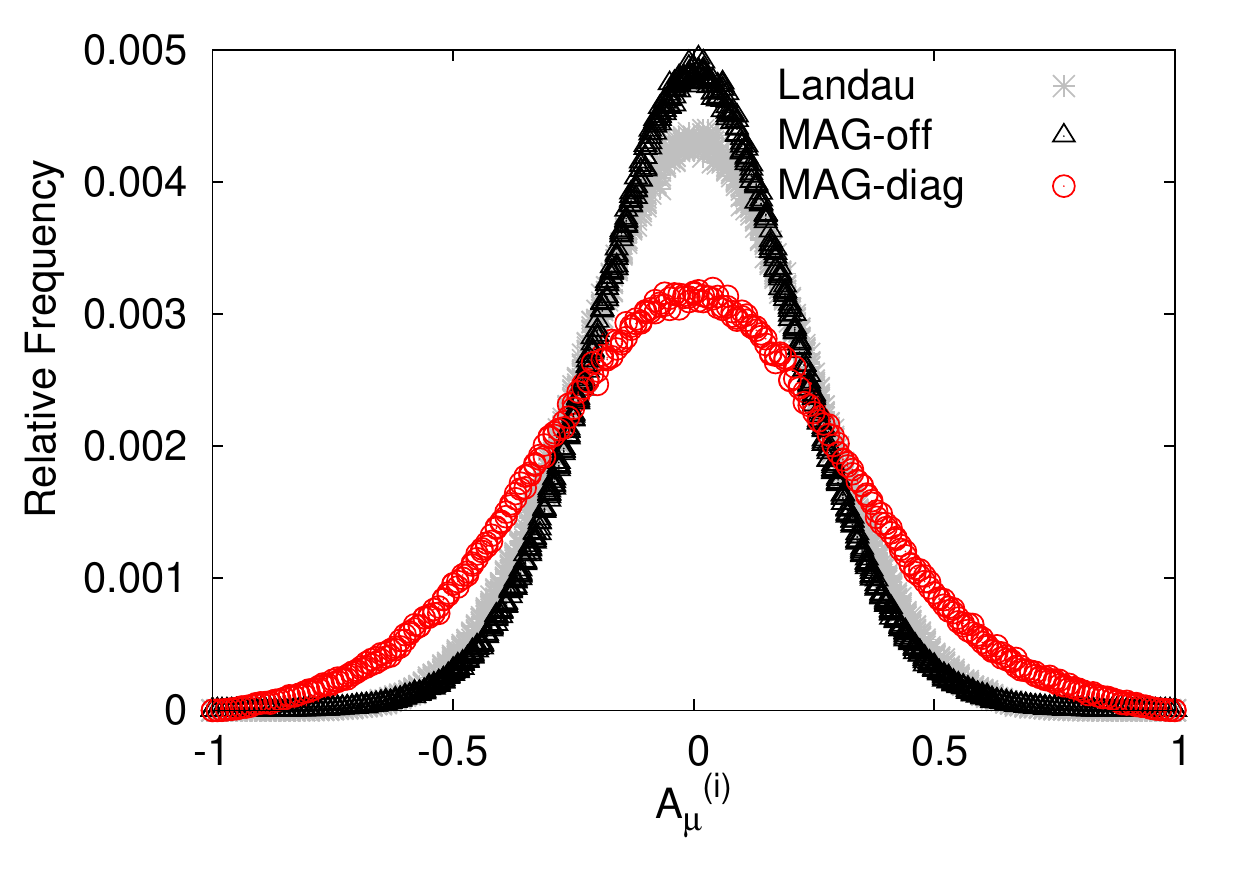}
	\caption{The relative frequency of the gluon field components $A_\mu^{(i)}(x)$ (in lattice units) from a single gauge field configuration in Landau gauge and MAG. In Landau gauge the eight field components follow the same normal distribution while in MAG a trend of the off-diagonal components (MAG-\emph{off}) towards smaller values compared to the diagonal components (MAG-\emph{diag}) is manifest.}
	\label{fig:histo}
\end{figure}

Once a maximum of \Eq{eq:MAG_SU3_functional} has been reached, the diagonal gluon fields, $A_\mu^{(i)}(x)$ with $i=3, 8$ are favoured over the off-diagonal gluons. In \fig{fig:histo} we show the distribution of the gluon fields $A_\mu^{(i)}(x)$ from a single lattice gauge field configuration. While in Landau gauge none of the gauge fields is favoured and thus their distribution lies on top of each other, the distribution of the MAG fields shows a clear shift of the diagonal fields towards larger values and correspondingly a shift towards smaller  values for the off-diagonal parts.

The MAG functional \Eq{eq:MAG_SU3_functional} is invariant under gauge transformations of the form
\be\label{eq:gd}
  g_d(x) = \exp\left(i\omega^{(3)}\lambda_3 + i\omega^{(8)}\lambda_8\right)
\ee
and therefore the MAG is an incomplete gauge condition; it leaves a remaining $U(1)_3\times U(1)_8$ gauge freedom. The 
latter we remove by enforcing that $U_\mu(x)$ fulfils in addition to the MAG the Landau gauge condition
\begin{equation}\label{eq:landau_functional}
	F_{\mathrm{Landau}}^{g}[U] = \Re\sum_{\mu, x} \tr{g(x)\,U_\mu(x)\,g(x+\hat\mu)^\dagger} \stackrel{!}{\longrightarrow}
\text{max.}\,,
\end{equation}
with respect to diagonal gauge transformations $g(x)=g_d(x)$.

Lastly, in order to study the diagonal and off-diagonal parts of the gluon fields separately, we extract the fields $A_\mu(x)$ from the link variables \Eq{eq:linkVars} via the first order approximation
\be\label{eq:linkToGaugeField}
	A_\mu(x) = \frac{1}{2iag_0}\left(U_\mu(x) - U_\mu(x)^\dagger\right)\Big|_{\mathrm{traceless}}\,.
\ee
Note that we do not make use of the ``exact'' logarithmic definition of the lattice gluon fields in order to stay 
consistent with the definition of the gluon fields in the Landau gauge condition \cite{Schrock:2012fj}.\footnote{We 
performed some checks using the logarithmic definition: the qualitative behavior of the gluon propagator is the same, 
as it was also found in Ref.~\cite{Ilgenfritz:2010gu}. The main difference here is, as expected, that the diagonal part 
of the longitudinal gluon propagator from the logarithmic definition does not vanish.}

\section{QCD propagators}\label{sec:QCDprops}

\subsection{Gluon propagator}
In Landau gauge the gluon propagator in momentum space is transverse and diagonal in color space
\begin{align}
  D_{\mu\nu}(k^2) &= \frac{1}{8} \sum_{i=1}^8 \left< A_\mu^{(i)}(k) A^{(i)}_\nu(-k) \right>\\
		  &= \left( \delta_{\mu\nu} - \frac{k_\mu k_\nu}{p^2} \right) D(k^2),
\end{align}
where $A_\mu^{(i)}(k)$ are the Fourier transformed gauge fields extracted from the links by 
Eq.~\eqref{eq:linkToGaugeField}. Appropriate for the Symanzik-improved L\"uscher--Weisz gauge 
action \cite{Bonnet:2001uh}, we define the momentum 
variable as
\be\label{eq:gluonmomenta}
	k_\mu = \frac{2}{a} \sqrt{ \sin^2\left(\frac{p_\mu a}{2}\right)+\frac{1}{3}\sin^4\left( \frac{p_\mu a}{2} 
\right) },
\ee
where
\be
p_\mu = \frac{2\pi n_\mu}{aL_\mu}
\ee
are the discrete lattice momenta.
The transversality of the momentum space propagator is a direct consequence of the Landau gauge condition $\partial_\mu 
A_\mu(x) = 0$.

For the MAG case 
we split the propagator in a diagonal
\be
  D^\text{diag}_{\mu\nu}(k^2) = \frac{1}{2} \sum_{i=3,8} \left< A_\mu^{(i)}(k) A^{(i)}_\nu(-k) \right>
\ee
and an off-diagonal part
\be
  D^\text{off}_{\mu\nu}(k^2) = \frac{1}{6} \sum_{i\neq3,8} \left< A_\mu^{(i)}(k) A^{(i)}_\nu(-k) \right>.
\ee
Due to the residual $U(1)_3\times U(1)_8$ Landau gauge fixing, the diagonal propagator is transverse, whereas the 
off-diagonal propagator has a longitudinal and a transverse component
\be
  D^\text{off}_{\mu\nu}(k^2) = \left( \delta_{\mu\nu} - \frac{k_\mu k_\nu}{k^2} \right) D^\text{off}_\text{T}(k^2) + 
 \frac{k_\mu k_\nu}{k^2} D^\text{off}_\text{L}(k^2).
\ee

\subsection{Quark propagator}
In manifestly covariant gauges, the interacting  quark propagator
$S(\mu;p^2)$, renormalized at the renormalization point $\mu$,
can be decomposed into Dirac scalar and vector parts
\be
	S(\mu;p^2) = \left(i\pslash A(\mu;p^2)+B(\mu;p^2)\right)^{-1}
\ee
or equivalently as
\be\label{fullcont}
	S(\mu;p^2) = Z(\mu;p^2)\left(i\pslash +M(p^2)\right)^{-1}.
\ee
In the last equation 
the wave-function renormalization function $Z(\mu;p^2)=1/A(\mu;p^2)$ carries all the information about the 
renormalization scale and
the mass function $M(p^2)=B(\mu;p^2)/A(\mu;p^2)$ is a renormalization group invariant.

The lattice regularized quark propagator
$S_L(p^2;a)$, which depends on the lattice spacing $a$, can then be renormalized at  renormalization scale $\mu$ 
with the momentum independent quark wave-function renormalization constant $Z_2(\mu;a)$,
\be
	S_L(p^2;a) = Z_2(\mu;a) S(\mu;p^2).
\ee
The momentum subtraction scheme (MOM) has the renormalization point boundary conditions
$Z(\mu;\mu^2)=1$ and $M(\mu^2)=m(\mu)$ where $m(\mu)$ is the running mass.

The nonperturbative functions $M(p^2)$ and 
$ Z(p^2) \equiv Z_2(\mu;a)Z(\mu;p^2)$ can be extracted
directly from the lattice.
To this end we invert the Asqtad fermion matrix \cite{Orginos:1999cr} in order to obtain the quark propagator which we subsequently Fourier transform to momentum space. Taking basic Clifford algebra properties into account we can extract the dressing functions. For details we refer to \cite{Bowman:2001xh, Bowman:2002bm}.
Note that the lattice dressing functions will be functions of the lattice quark momenta (which differ from the gluon momenta \eq{eq:gluonmomenta}) and for the Asqtad action these are defined by
\be
	k_\mu = \sin(p_\mu)\left( 1 + \frac{1}{6}\sin^2(p_\mu) \right).
\ee
We perform a cylinder-cut \cite{Skullerud:2000un} on all our data and average
over the discrete rotational and parity symmetries of $S_L(p^2;a)$
to increase statistics.

\section{Results}\label{sec:results}
\subsection{Gauge configurations}
For our simulation we adopted two sets of gauge field 
configurations generated by the MILC collaboration \cite{Bernard:2001av, Aubin:2004wf, Aubin:2009jh, Bazavov:2009bb}:
a ``coarse'' set of size $20^3\times 64$ with lattice spacing $a=\unit[0.12]{fm}$, which consists of five dynamical plus a quenched ensemble, and furthermore a ``fine'' set consisting of a single ensemble of size $40^3\times 96$ with lattice spacing $a=\unit[0.09]{fm}$. 
% and analyzed using mostly the C++ toolkit \texttt{FermiQCD} \cite{DiPierro:2003sz}.
The configurations were generated with the Symanzik-improved L\"uscher--Weisz 
gauge action \cite{Luscher:1984xn} and have been made available to the lattice community 
via the Gauge Connection \cite{DiPierro:2011aa}. Both sets include two light 
degenerate ($l$) and one heavier quark flavor ($s$) (except the quenched ensemble), implemented with the Asqtad improved 
action \cite{Orginos:1999cr}. The parameters of the lattices are summarized in 
Table~\ref{tab:setup}; for the reported lattice scales 
and quark masses we refer to the original work \cite{Bernard:2001av, Aubin:2004wf, Aubin:2009jh, Bazavov:2009bb}.

\begin{table}[ht]
\begin{center}
\begin{tabular}{|c|c|c|c|c|}
\hline
 $N_s^3\times N_t$ & $a\,[\mathrm{fm}]$ & $m_l\,[\mathrm{MeV}]$ & $m_s\,[\mathrm{MeV}]$ & \# configs.\\\hline\hline
 \multirow{5}{*}{$20^3\times 64$} & \multirow{5}{*}{0.12} & 11.5 & \multirow{5}{*}{82.2} & 976\\
 & & 16.4 & & 573\\
 & & 32.9 & & 391\\
 & & 49.3 & & 432\\
 & & 65.8 & & 350\\
 & & $\infty$ & $\infty$ & 408\\\hline
 $40^3\times 96$ & 0.09 & 6.8  & 68.0 & 187 \\\hline
\end{tabular}
\end{center}
\caption{Overview of the gaugefield parameters: the lattice size $N_s^3\times N_t$, lattice spacing $a$, dynamical quark masses $m_l$ and $m_s$ ($\infty$ indicating quenched gauge fields) and the number of configurations that enter our analysis. 
}
\label{tab:setup}
\end{table}

\subsection{Gluon propagator}
\begin{figure*}[htb]
	\center
	\includegraphics[width=.99\columnwidth]{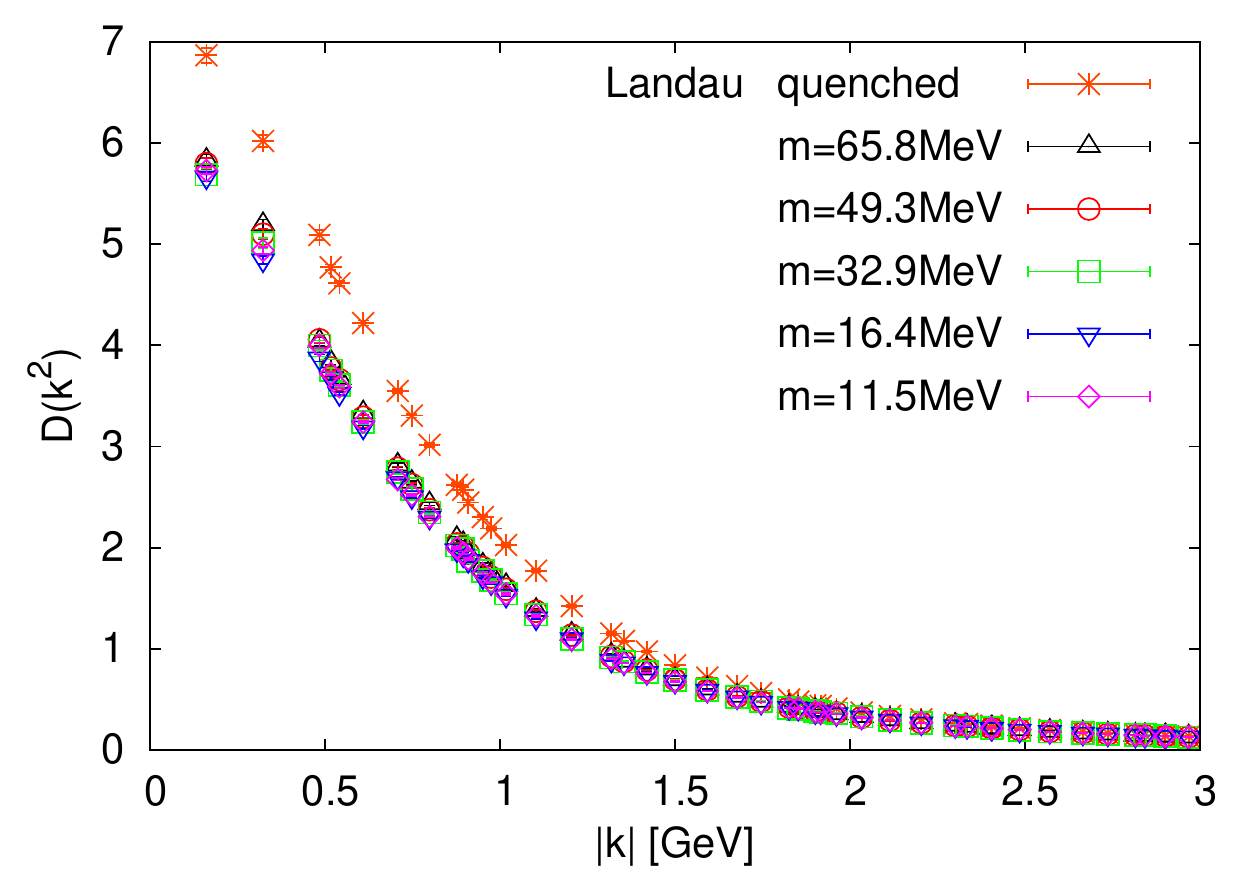}
	\includegraphics[width=.99\columnwidth]{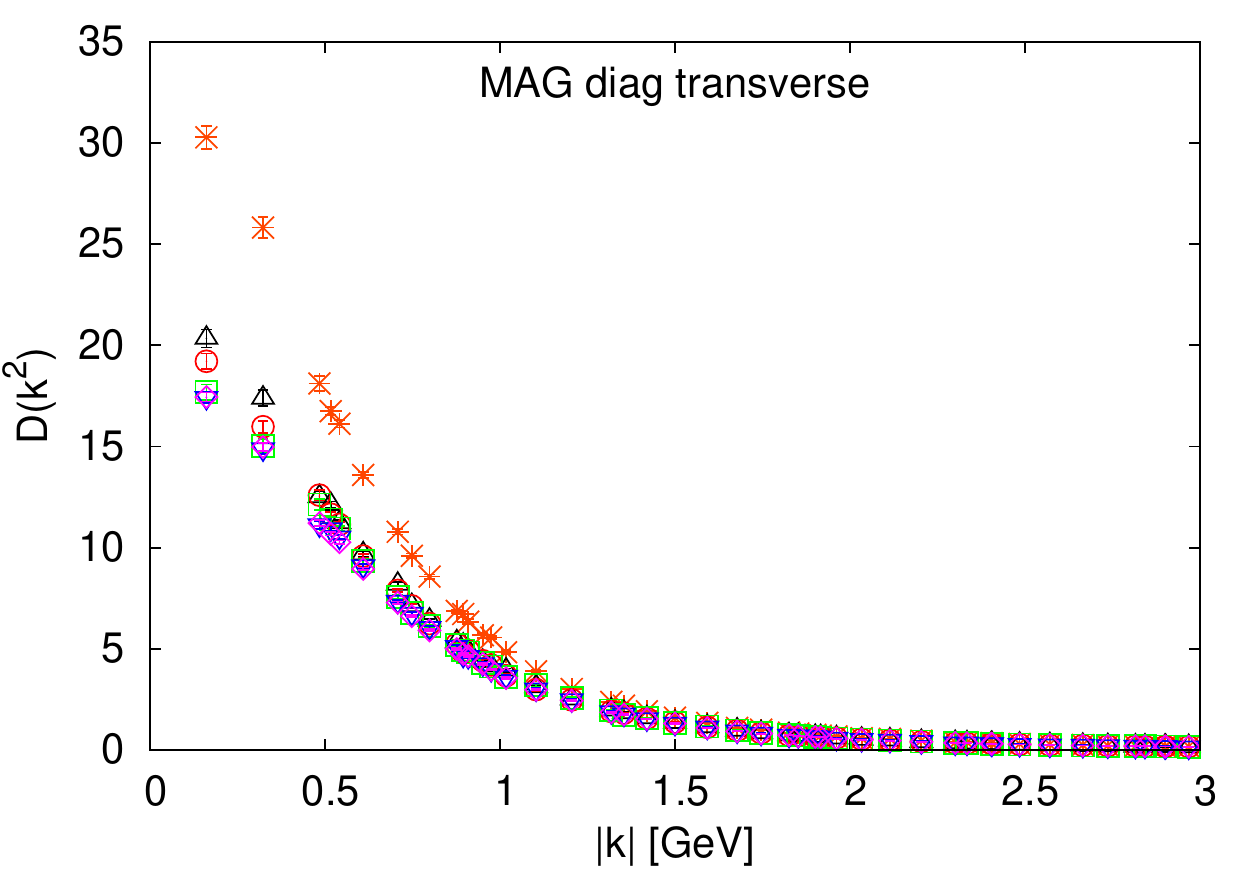}\\
	\includegraphics[width=.99\columnwidth]{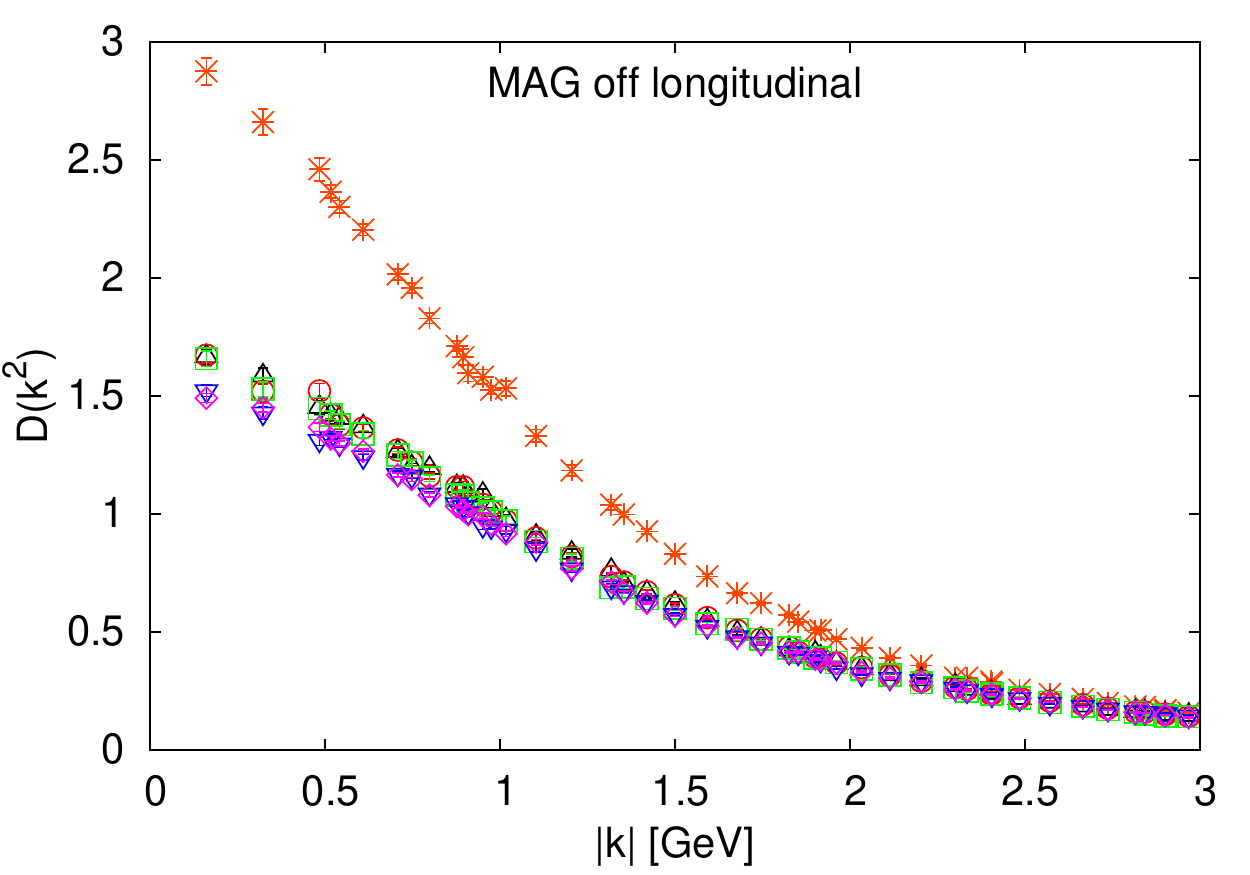}
	\includegraphics[width=.99\columnwidth]{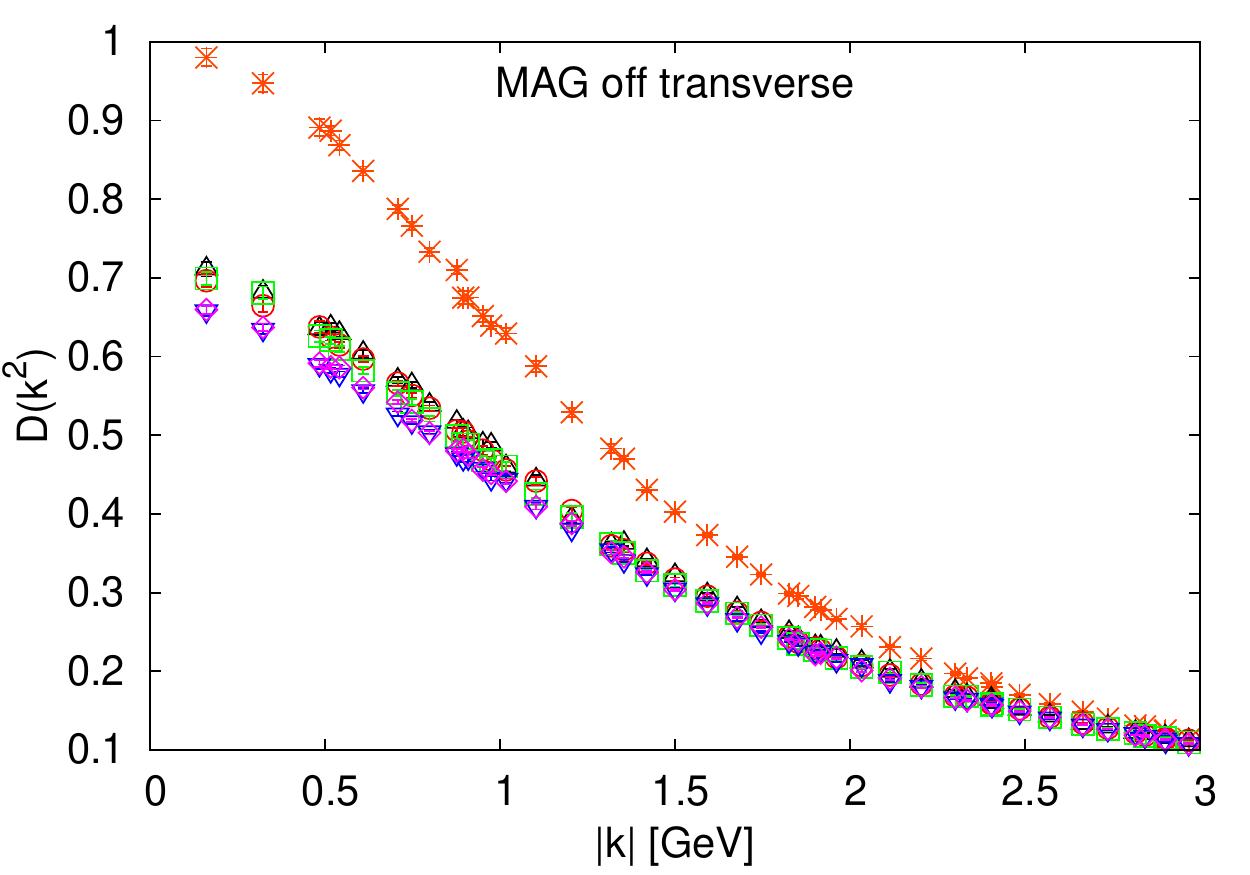}
	\caption{The gluon propagator of the coarse ensembles of \Tab{tab:setup} renormalized at $\mu = \unit[4]{GeV}$.}
	\label{fig:gluon_2064f21}
\end{figure*}

The gluon propagator in the maximally Abelian gauge has already been studied both in SU(2) \cite{Amemiya:1998jz, 
Bornyakov:2003ee} and SU(3) \cite{Gongyo:2012jb, 
Gongyo:2013sha} for pure Yang--Mills theory.
Here we extend those studies to full QCD with $2+1$ flavors of dynamical quarks.
In Fig.~\ref{fig:gluon_2064f21} we show the propagators of the coarse ensemble for the 
different quark masses including the quenched dataset. In addition to the three MAG propagators (diagonal transverse, 
off-diagonal longitudinal and off-diagonal transverse) we include the Landau gauge propagator for comparison. The 
Landau gauge propagator has been studied on the same dataset in \cite{Bowman:2004jm}. Following their setup we 
use the same renormalization condition
\be
D(k^2 = \mu^2) = \frac{1}{\mu^2}
\ee
at $\mu = \unit[4]{GeV}$.
In the IR we find suppression of the propagator with dynamical quarks due to screening effects. Compared to Landau gauge, this effect is more pronounced in the 
maximally Abelian gauge. Decreasing the quark mass leads to a 
further suppression in the IR, however the dependence on the quark mass is small.

\begin{figure}[htb]
	\center
	\includegraphics[width=.99\columnwidth]{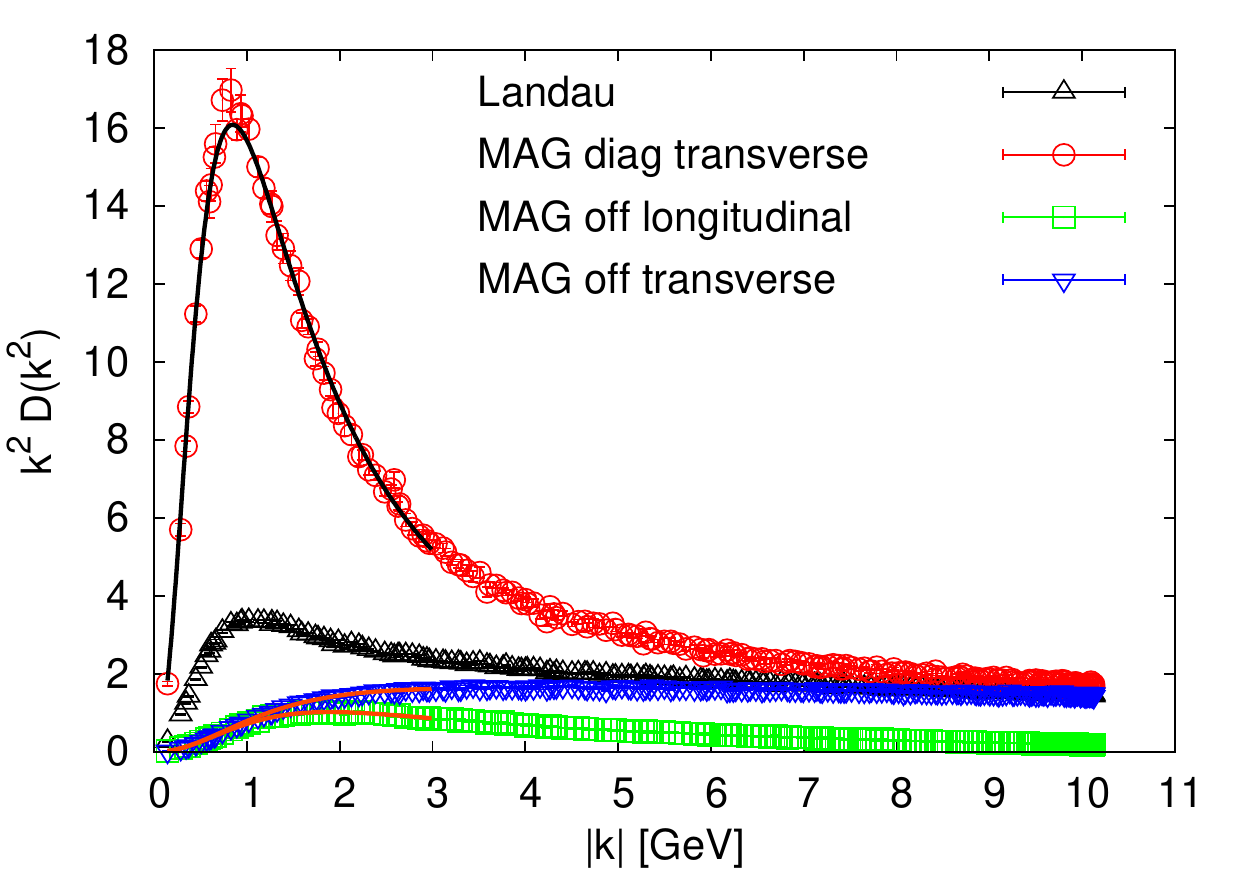}
	\caption{The gluon form factors of the fine ensemble. The solid lines illustrate the fit to 
Eq.~\eqref{eq:gluon_fit_suganuma}.}
	\label{fig:gluon_4096f21}
\end{figure}

In Fig.~\ref{fig:gluon_4096f21} we present the gluon form factors of the fine ensemble.
It is very important to note that in both Figs.~\ref{fig:gluon_2064f21} and \ref{fig:gluon_4096f21}, it is obvious that the diagonal parts of the MAG gluon propagator are pronounced as compared to the Landau gauge counterparts, and, respectively, the off-diagonal parts are suppressed. This extends previous findings of IR Abelian dominance from pure Yang--Mills theory to full QCD.

Our findings for the fine ensemble in Fig.~\ref{fig:gluon_4096f21} are in very good agreement with the SU(2) results 
of \cite{Bornyakov:2003ee} considering the fact that we study the SU(3) propagator and included dynamical quarks. 
There the authors found a wide maximum of the dressing function $k^2 D^\text{off}_\text{L}(k^2)$ at around 2 GeV 
and a sharp peak of $k^2 D^\text{diag}(k^2)$ around 0.7 GeV.
To compare our results with the quenched SU(3) results of \cite{Gongyo:2013sha} we applied a fit to our data with the 
their function
\be
\label{eq:gluon_fit_suganuma}
  D(k^2) = \frac{Z}{(k^2+m^2)^\nu}
\ee
in the same momentum regime $k < \unit[3]{GeV}$. Qualitatively, the results compare well to the quenched results from 
Ref.~\cite{Gongyo:2013sha}.

\begin{table}[ht]
\begin{center}
\begin{tabular}{|l|c|c|c|c|}
\hline
  & $m$ [GeV] & $\nu$ & Z & $\chi^2/\text{n.d.f.}$\\\hline\hline
 $D^\text{off}_\text{T}$ & 1.47(2)& 1.18(2) & 14.9(6) & 0.9\\
 $D^\text{off}_\text{L}$ & 1.66(3)& 1.77(4) & 36.2(36) & 1.0\\
 $D^\text{diag}$& 0.78(1) & 1.85(1) & 181.6(54) & 1.7\\\hline
 \end{tabular}
\end{center}
\caption{Results of a fit of the maximally Abelian gluon propagators to \eqref{eq:gluon_fit_suganuma}.}
\label{tab:gluonpropagator}
\end{table}

\subsection{Quark propagator}
We obtain the quark propagator in the standard way by inverting the (Asqtad) Dirac operator for a point source on a gauge field background. This is performed in Landau gauge and in the maximally Abelian gauge with $U(1)_3\times U(1)_8$ Landau residual gauge fixing. Additionally, we split the maximally Abelian gauge gluon fields in their diagonal and off-diagonal components and invert the Dirac operator on the two parts separately. Thus, we obtain four ``kinds'' of quark propagators from each ensemble.

\subsubsection{Quark mass dependence}
On the five dynamical coarse ensembles of \Tab{tab:setup}, we calculate the quark propagator in Landau gauge, in 
maximally Abelian gauge (MAG), on a pure diagonal MAG background (MAG-\emph{diag}) and a pure off-diagonal MAG 
background (MAG-\emph{off}). The mass parameter of the valence quark propagator has been set to the value of the 
corresponding light sea quark mass for these five ensembles. 
This will allow for a systematic extrapolation to the chiral limit.

\begin{figure*}[htb]
	\center
	\includegraphics[width=0.99\columnwidth]{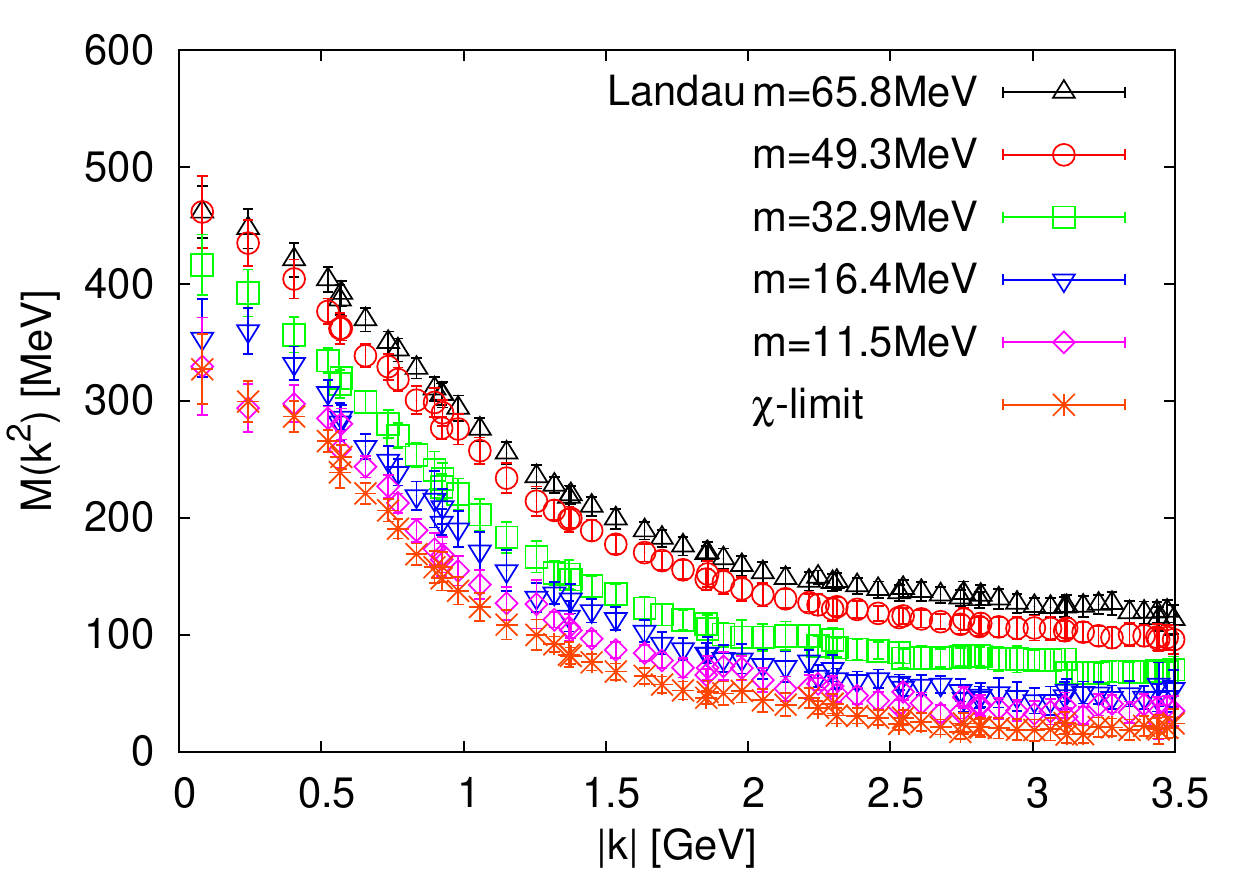}
	\includegraphics[width=0.99\columnwidth]{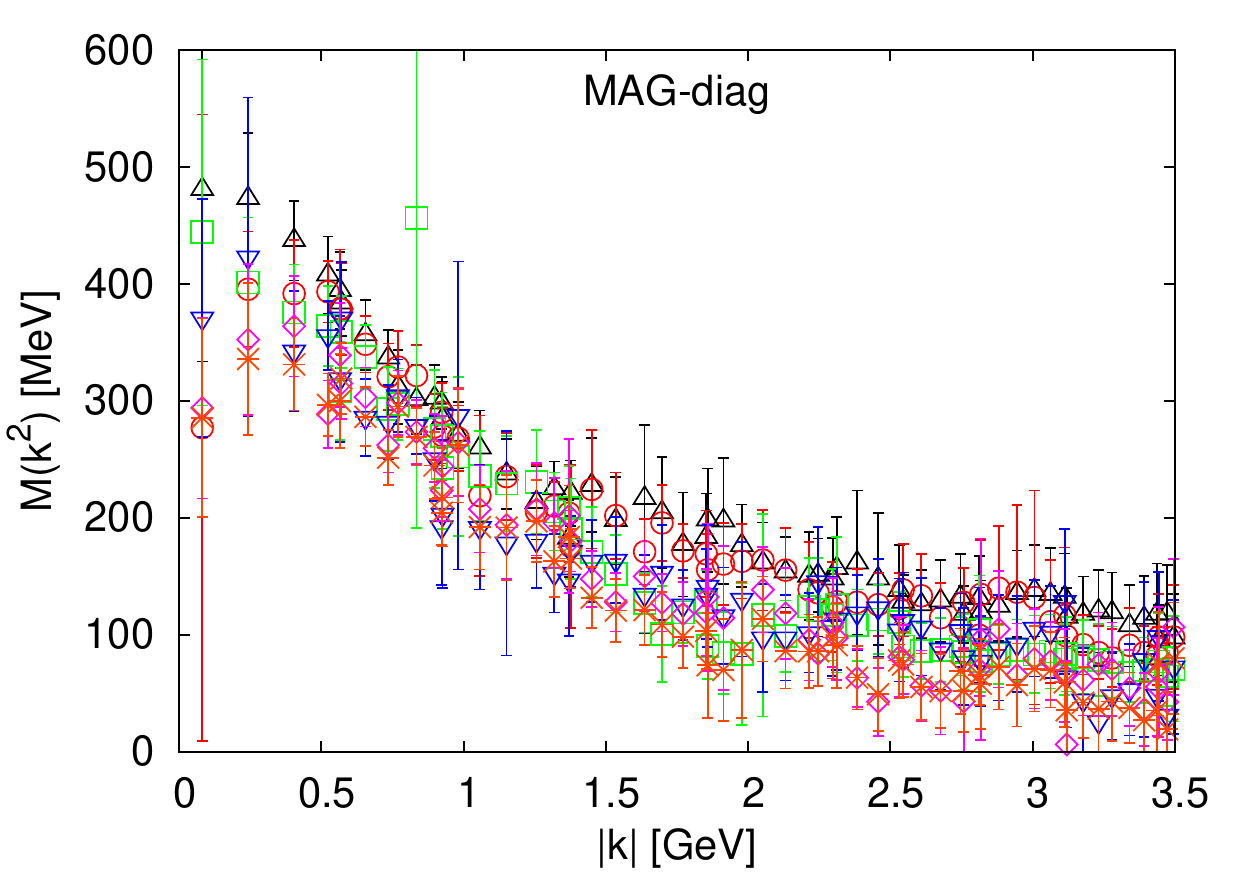}\\
	\includegraphics[width=0.99\columnwidth]{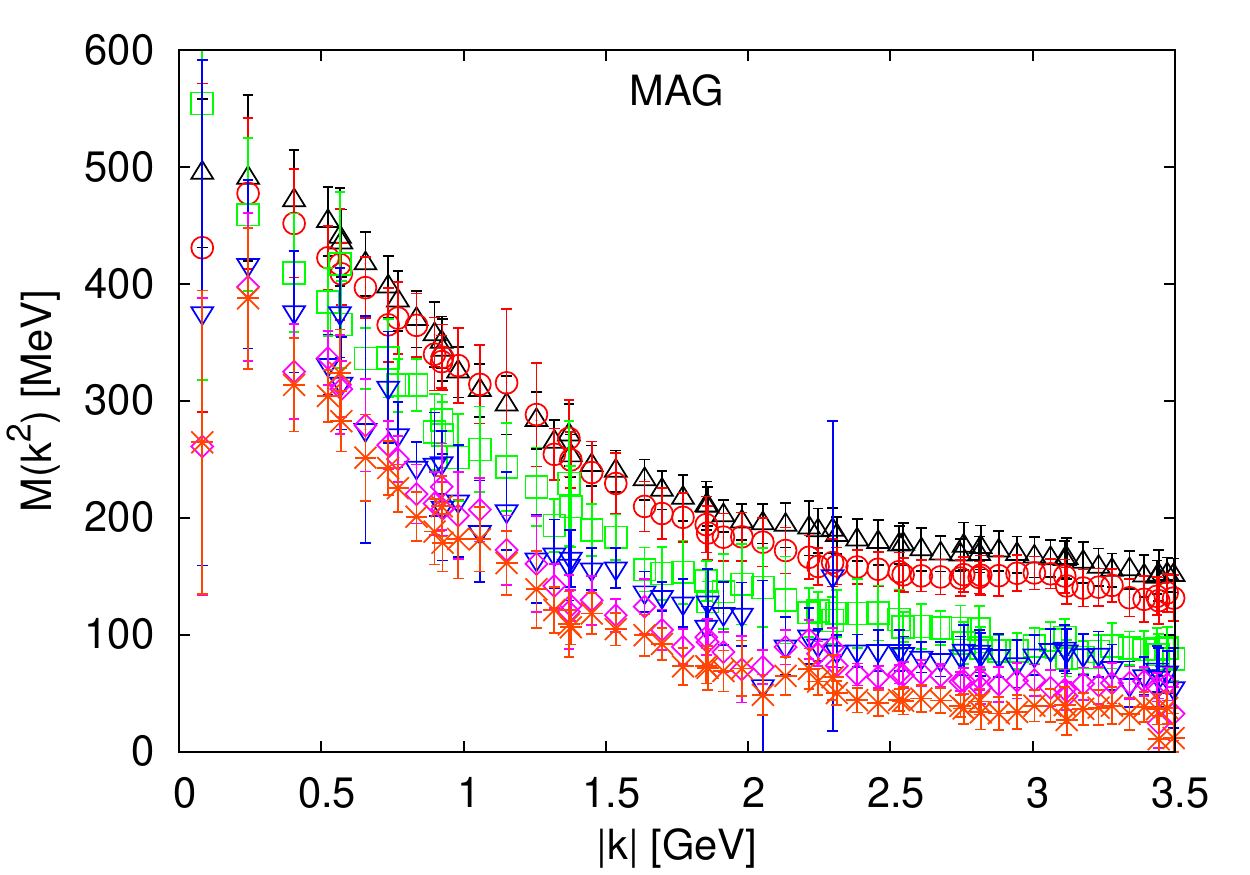}
	\includegraphics[width=0.99\columnwidth]{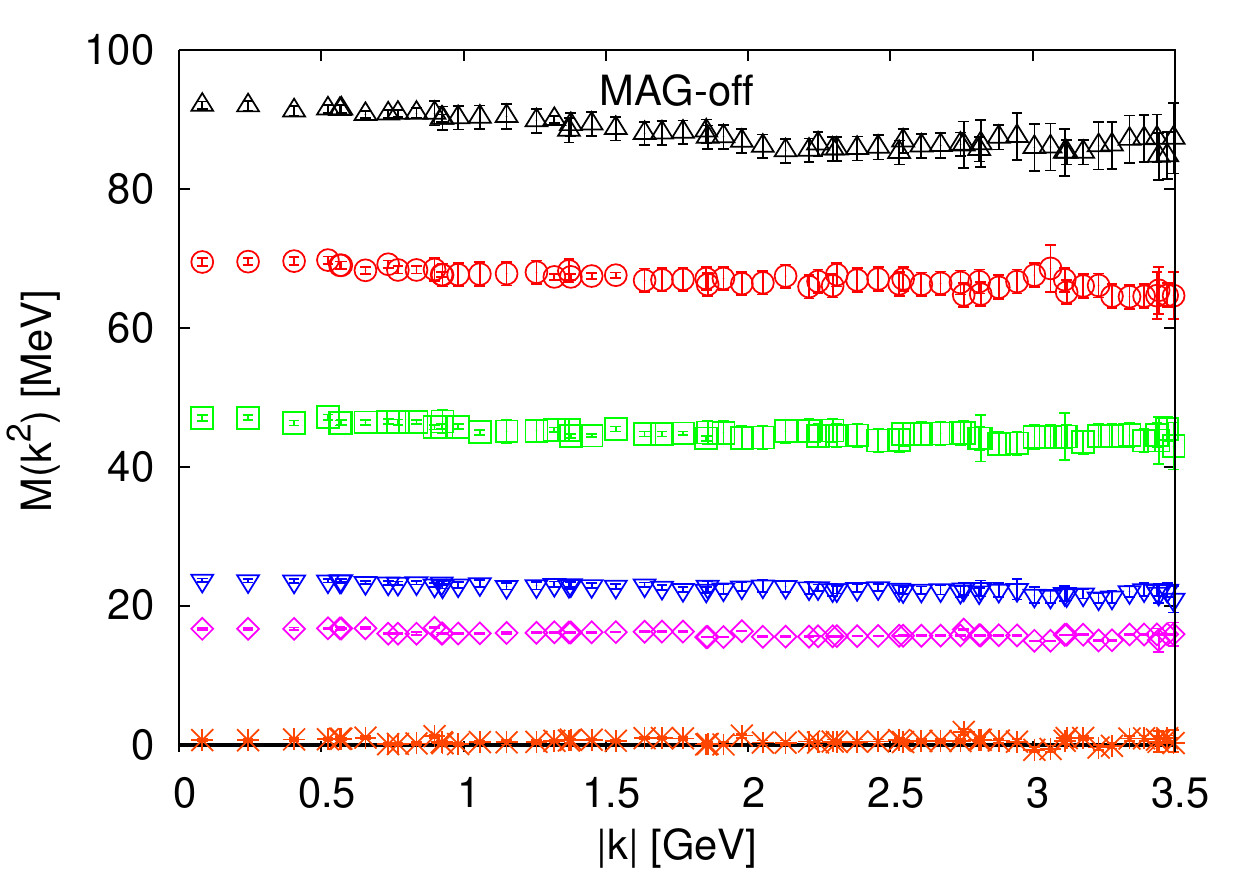}
	\caption{The quark mass function from the coarse ensembles of \Tab{tab:setup}. The bare quark masses are set to the values of the dynamical light quark masses. Additionally, a linear extrapolation to the chiral limit is shown.}
	\label{fig:M_2064f21}
\end{figure*}

\begin{figure*}[htb]
	\center
	\includegraphics[width=0.99\columnwidth]{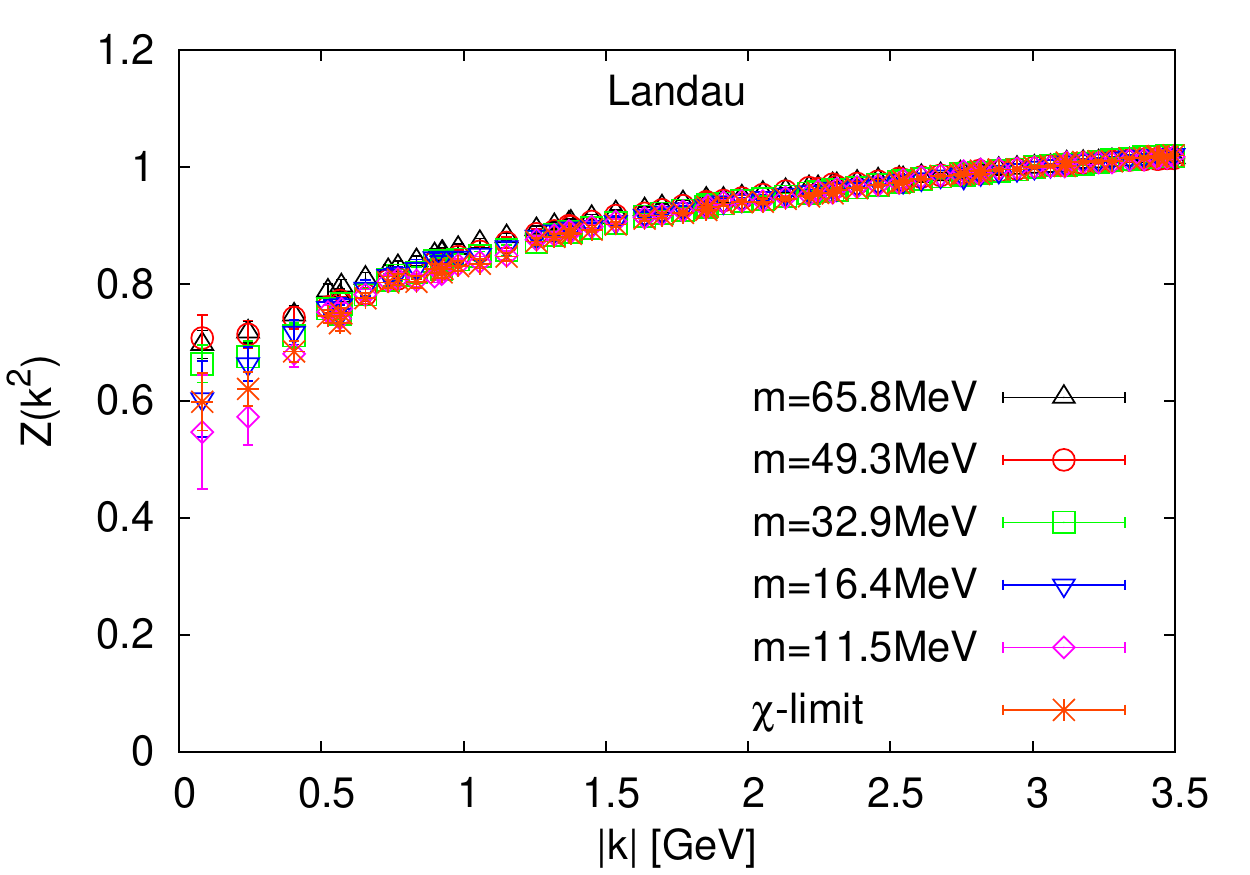}
	\includegraphics[width=0.99\columnwidth]{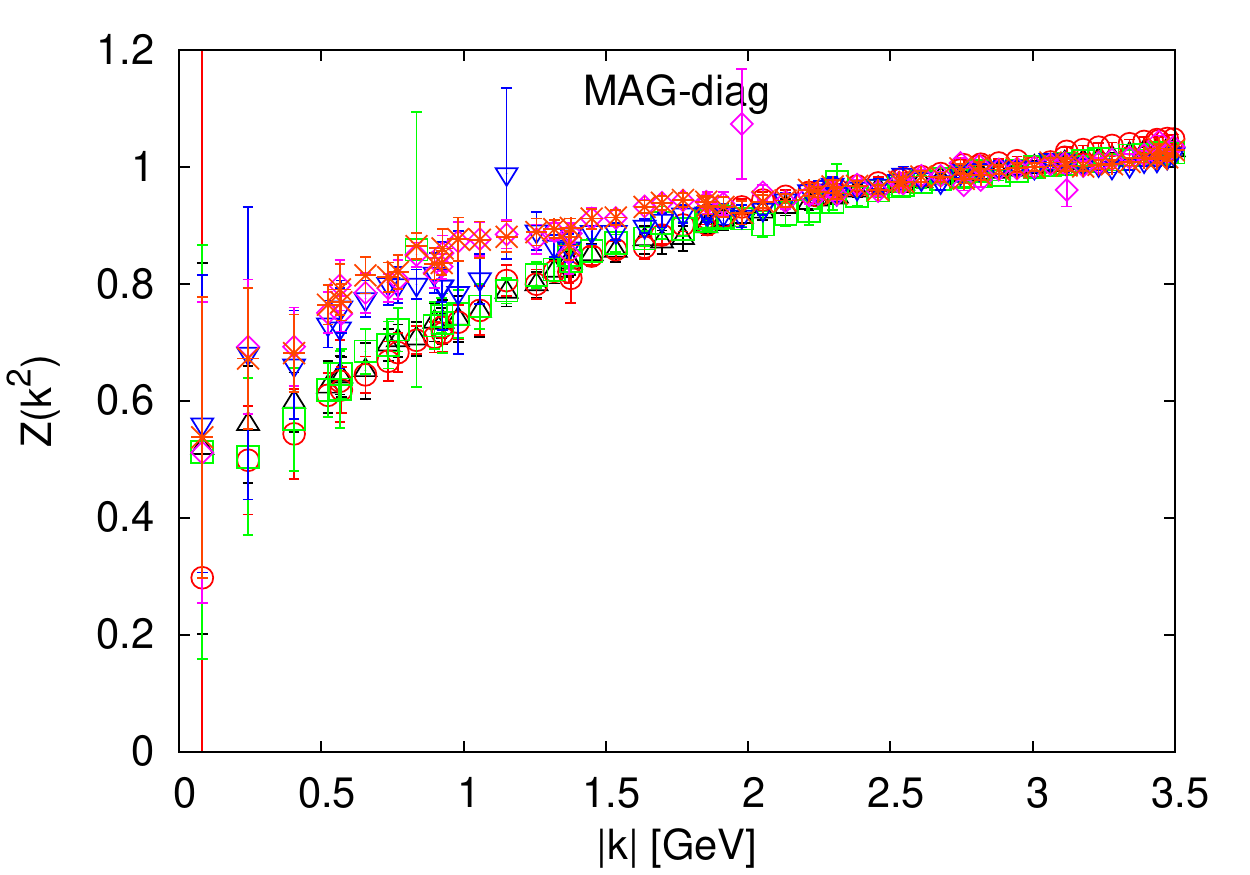}\\
	\includegraphics[width=0.99\columnwidth]{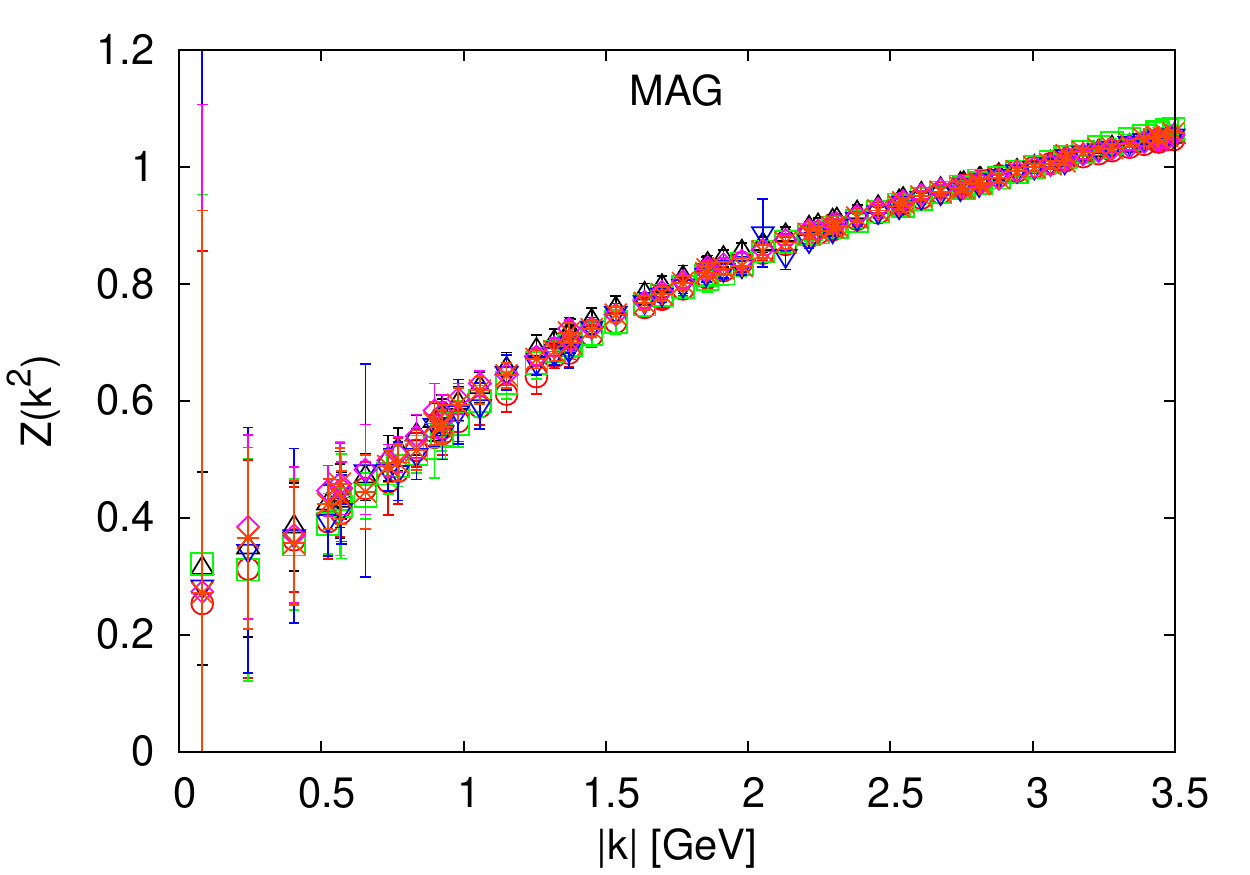}
	\includegraphics[width=0.99\columnwidth]{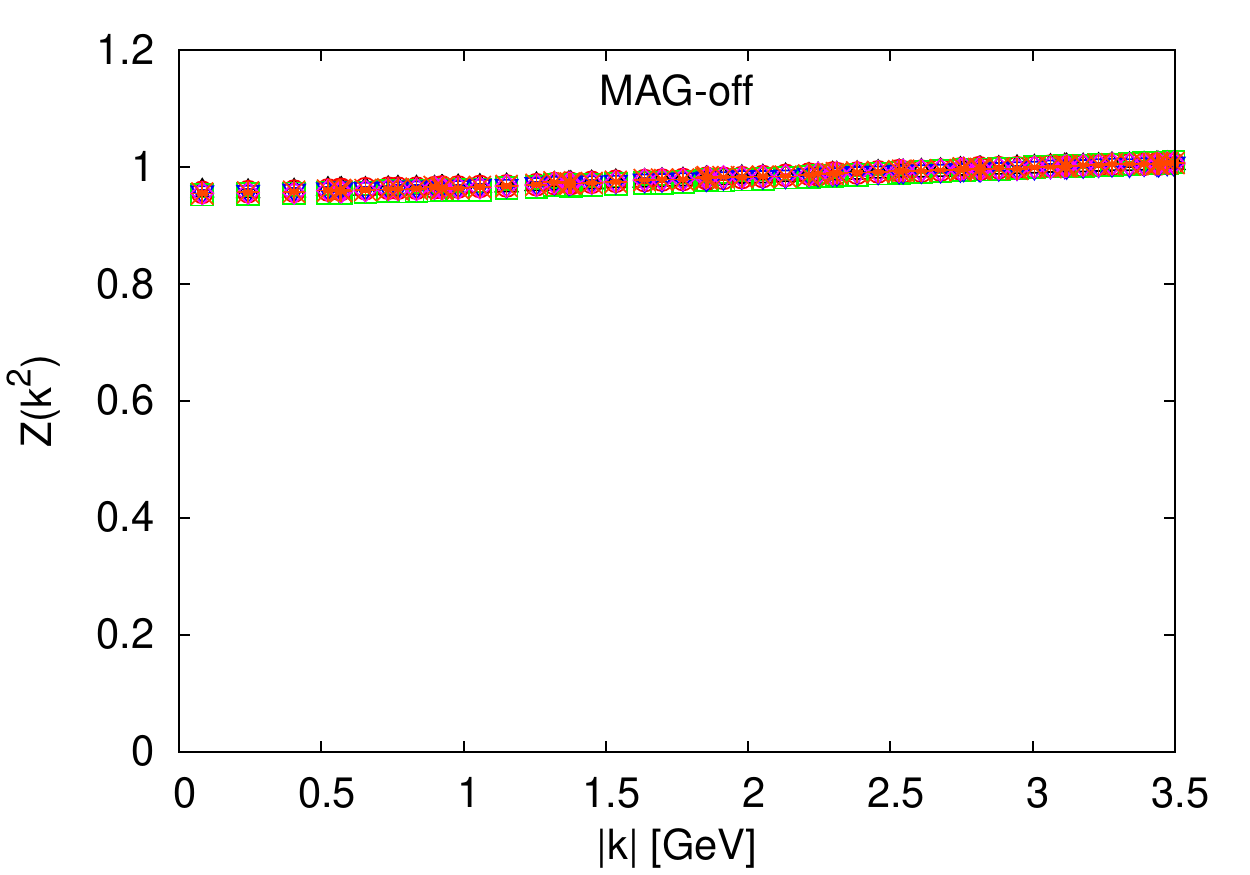}
	\caption{The quark function renormalization function from the coarse ensembles of \Tab{tab:setup}. The bare quark masses are set to the values of the dynamical light quark masses. A linear extrapolation to the chiral limit is shown. All data have been been renormalized at $\mu=\unit[3]{GeV}$.}
	\label{fig:Z_2064f21}
\end{figure*}

In \fig{fig:M_2064f21} the quark mass function $M(k^2)$ is shown for the four types of gluon backgrounds (Landau gauge, MAG, MAG-\emph{diag} and MAG-\emph{off}) from all coarse ensembles, including a linear extrapolation to the chiral limit. In analog, \fig{fig:Z_2064f21} shows the corresponding quark wave-function renormalization functions $Z(k^2)$.

When comparing the Landau gauge quark propagator to the quark propagator in  MAG, the first observation is that the MAG data, with the same statistics, results in more gauge noise from the Monte Carlo integration. 
Moreover, the running masses of the MAG data lie higher than the corresponding Landau gauge masses, which holds from the largest mass of $m_l=\unit[65.8]{MeV}$ down to the chiral limit. The dynamically generated infrared masses, on the other hand, appear compatible within the error bars.

It is evident that $M(k^2)$ from the MAG-\emph{diag} gluon background nicely  resembles the Landau gauge analog (despite 
being more noisy), whereas $M(k^2)$ from MAG-\emph{off} gluons is constant, lying roughly $40\%$ higher than the corresponding bare quark mass. In the chiral limit 
it is compatible with zero. Similarly, the wave-function renormalization function  from MAG-\emph{off} gluons 
comes out close to its tree-level value, $Z(k^2)\approx 1$, independent of the quark mass.

%The MAG and MAG-\emph{off} renormalization function are stronger IR suppressed as compared to Landau gauge and MAG-\emph{diag}. The latter is compatible with Landau gauge, noticeably especially towards the chiral limit.

\subsubsection{Infrared behavior}

\begin{figure*}[htb]
	\center
	\includegraphics[width=0.99\columnwidth]{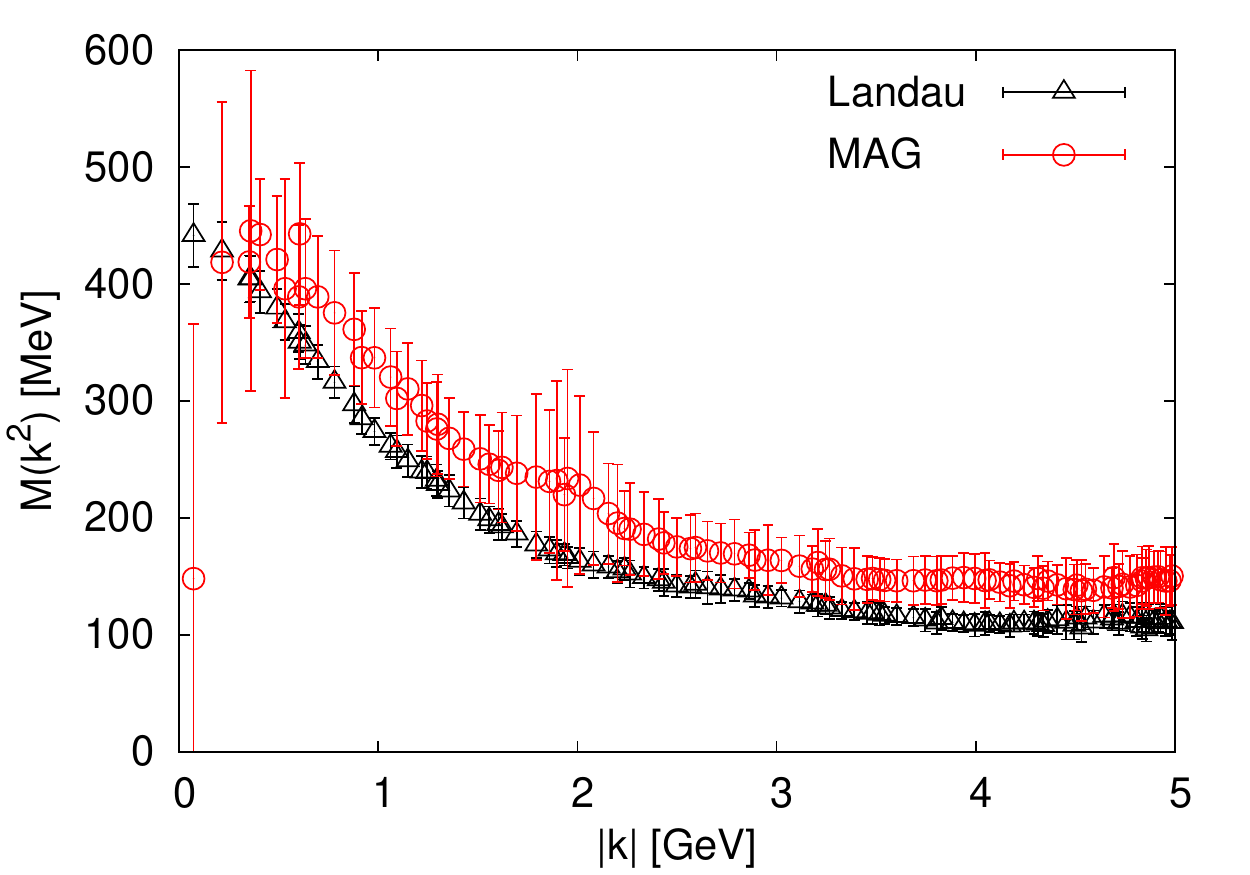}	 
	\includegraphics[width=0.99\columnwidth]{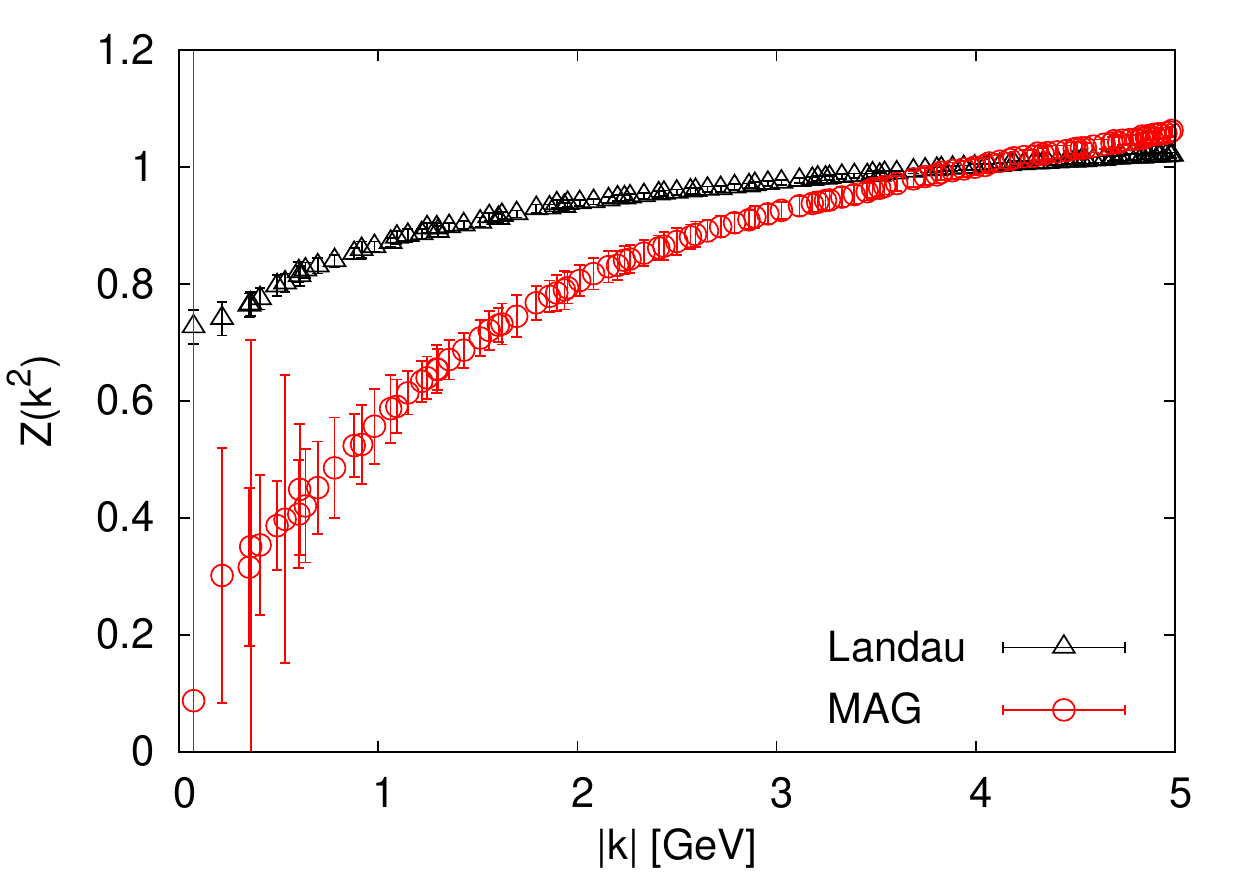}\\
	\includegraphics[width=0.99\columnwidth]{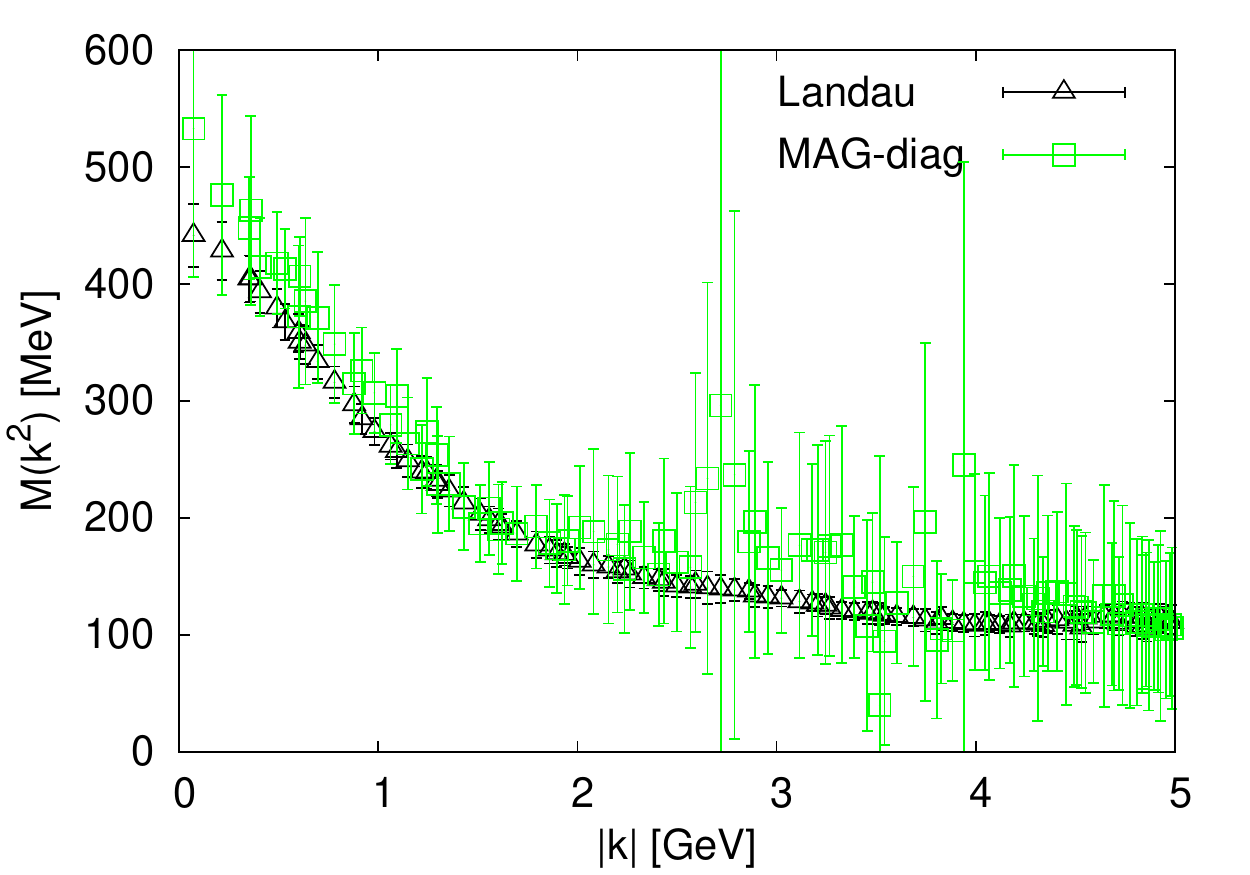}
	\includegraphics[width=0.99\columnwidth]{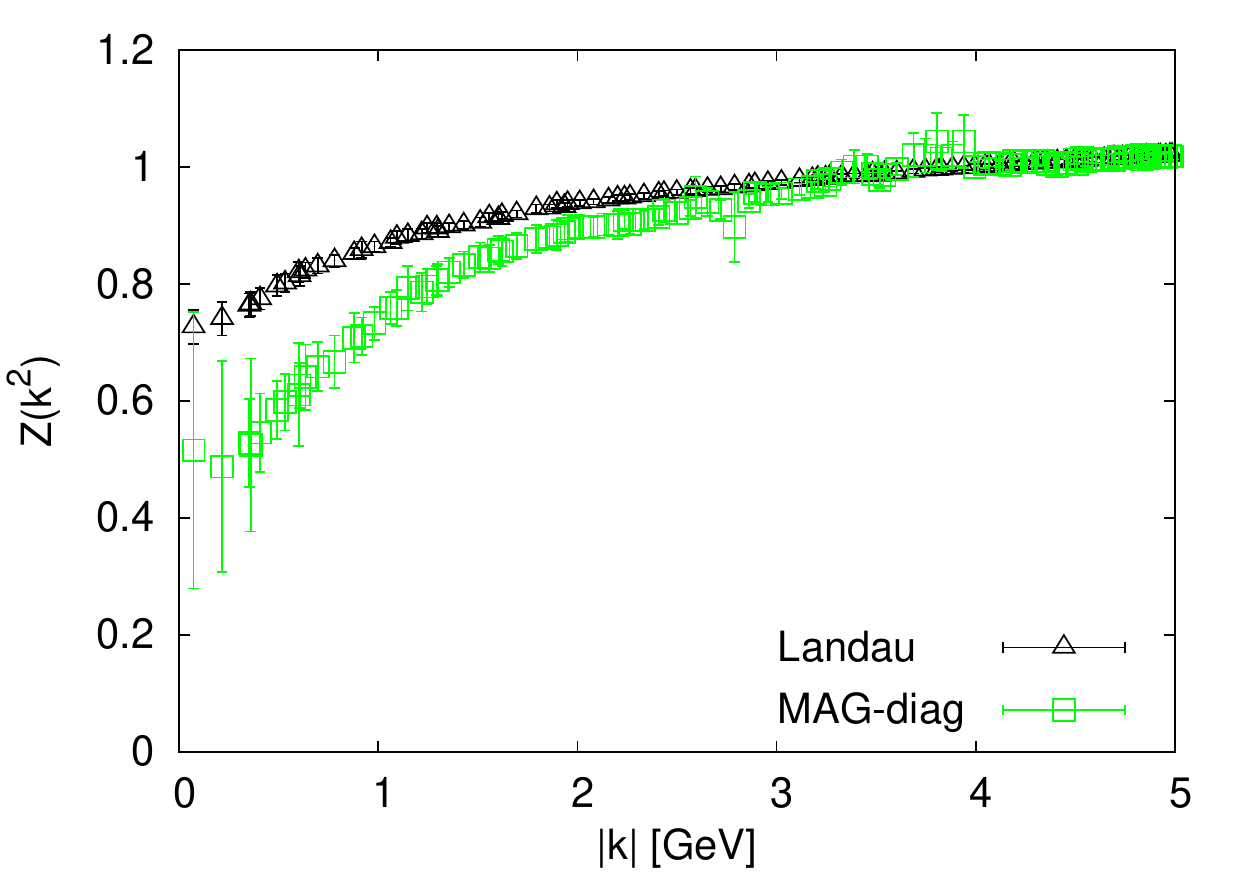}\\
	\includegraphics[width=0.99\columnwidth]{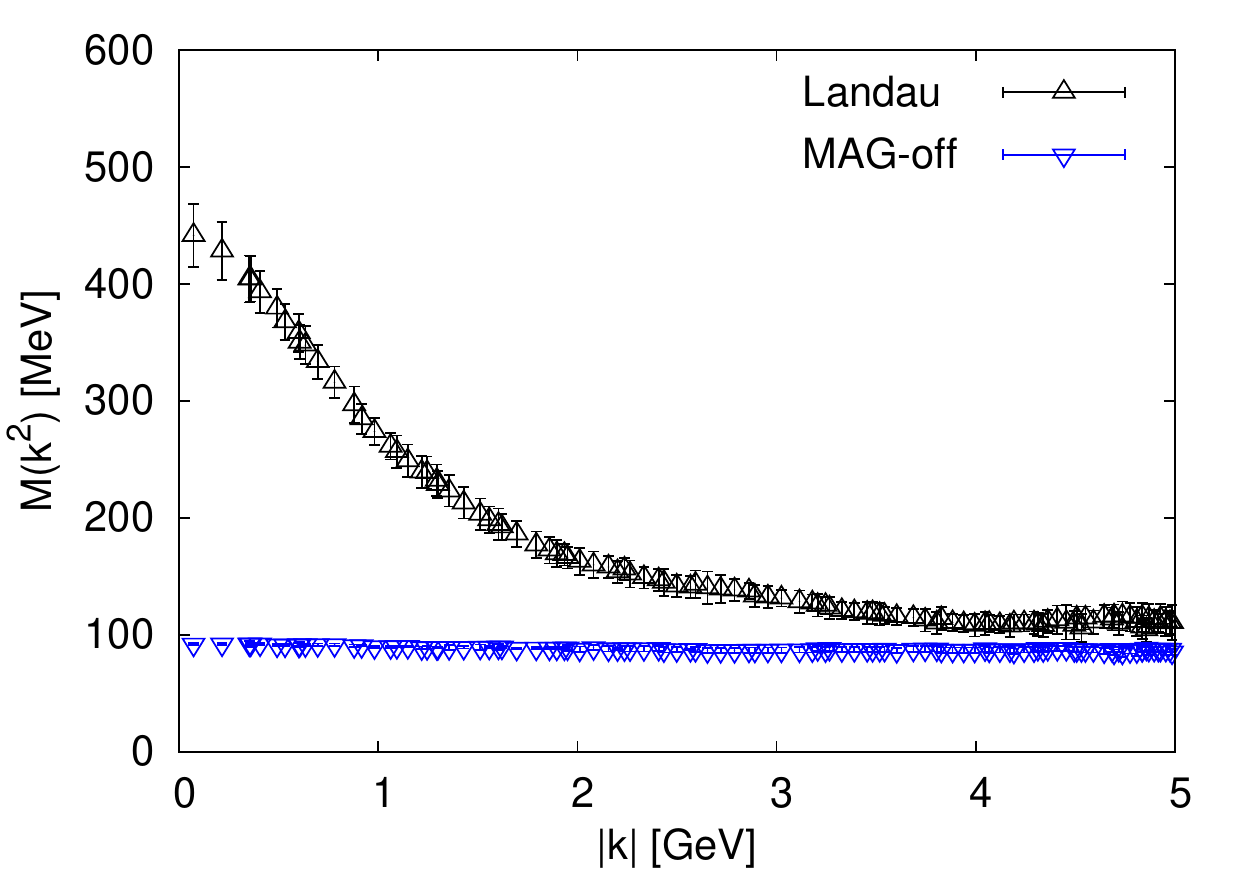}
	\includegraphics[width=0.99\columnwidth]{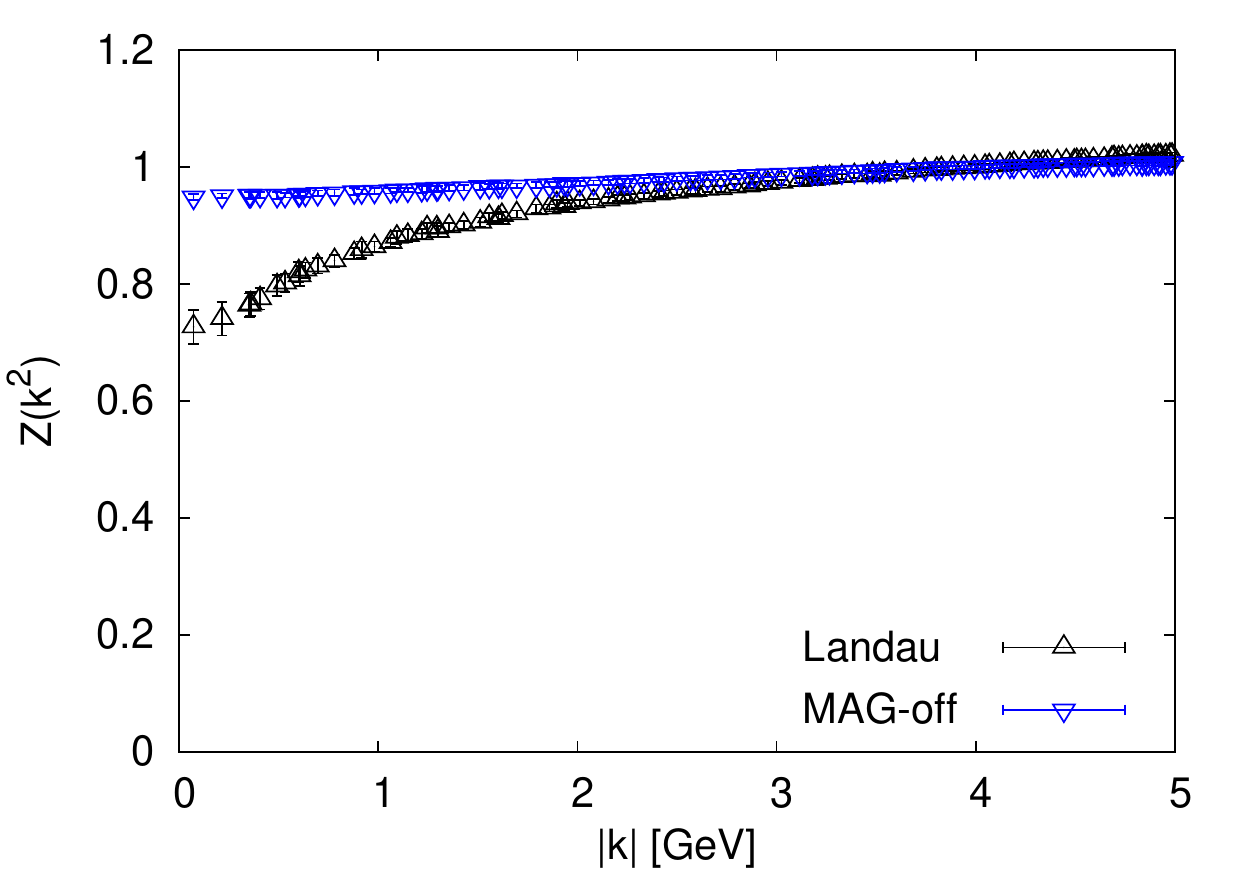}
	\caption{The quark mass function (left column) and renormalization function (right column) from the fine ensemble of \Tab{tab:setup}. The MAG (top row), MAG diagonal part (middle row) and MAG off-diagonal part (bottom row) are shown. Each of the three rows of plots includes additionally the Landau gauge quark propagator for direct comparison. The quark mass is equal to the dynamical strange quark mass ($m_s=\unit[68.0]{MeV}$). $Z(k^2)$ has been renormalized at $\mu=\unit[4]{GeV}$.}
	\label{fig:MZ_4096f21}
\end{figure*}

The coarse ensembles cannot provide a clear picture of the infrared behavior of the quark propagators. To improve thereon we adopt the fine MILC ensemble of \Tab{tab:setup}.
In order to keep the gauge noise and the simulation costs at an acceptable level, we use a valence quark with the 
mass of the heavier dyamical strange quark, $m_s=\unit[68.0]{MeV}$, instead of setting it to the light quark mass of 
$m_s=\unit[6.8]{MeV}$.

In \fig{fig:MZ_4096f21} we compare $M(k^2)$ and $Z(k^2)$ in MAG, MAG-\emph{diag} and MAG-\emph{off}, respectively, directly to the Landau gauge counterparts.
While the MAG and Landau gauge mass functions agree within the error bars over the whole momentum range, 
%$M(k^2) = B(k^2)/A(k^2)$ in MAG shows an abrupt drop in the IR around $\sim\unit[0.5]{GeV}$. In accordance therewith, 
$Z(k^2)$ appears to be stronger IR suppressed in MAG as compared to Landau gauge. 
%This should be investigated in more detail with larger lattices and larger statistics in a future study.

Similarly, the MAG-\emph{diag} mass function agrees over the whole momentum range within the error bars with 
its Landau gauge counterpart. The renormalization function exhibits qualitatively the same IR behavior as in 
Landau gauge. 
In contrast, the MAG-\emph{off} quark dressing functions hardly show any non-trivial dynamics.
This is the main finding of this study:
it shows that the Abelian parts of the gluon fields not only dominate the purely gluonic interactions, but also the infrared interactions of quarks. 

\section{Summary}\label{sec:summary}
We have fixed dynamical SU(3) lattice gauge fields to the combined maximally Abelian gauge and $U(1)_3\times U(1)_8$ Landau gauge.
From the lattice link variables we have extracted the continuum gluon fields which we subsequently separated into purely diagonal (Abelian) and off-diagonal components. 

We investigated the gluon propagator from diagonal and off-diagonal gluon fields. Dynamical quarks 
lead to an IR suppression compared to the quenched case. The suppression becomes stronger when decreasing the 
quark mass. The screening is more pronounced in the MAG propagators compared to the Landau gauge 
propagator. Our 
findings confirm the manifestation of infrared Abelian dominance in the gluon propagator as found in earlier studies on quenched 
lattices.

Finally, for the first time the maximally Abelian gauge quark propagator has been analyzed on a background of purely diagonal gluons as well as on the remaining, off-diagonal gluon background.
The hypothesis of Abelian dominance implies that the non-Abelian gluon field does not propagate at a long-distance scale and hence that only the Abelian component is relevant at a long-distance scale.
In accordance therewith, we have demonstrated that 
the quark propagator from a non-Abelian gluon background hardly shows any effects while
the quark propagator from an Abelian gluon 
background closely resembles its Landau gauge counterpart.

\begin{acknowledgments}
We would like to thank G.~Burgio, G.~'t~Hooft, V.~Mader and M.~Quandt for very valuable discussions. 
Special thanks go to M.~Quandt for providing constructive criticism on the manuscript.
\end{acknowledgments}

%--------------
% Bibliography:
%--------------
%\bibliographystyle{apsrev4-1}
%\bibliography{dissbib}


\end{document}